\documentclass[10pt, conference]{IEEEtran}
\IEEEoverridecommandlockouts

\usepackage{cite}
\usepackage{amsmath,amssymb,amsfonts}
\usepackage[normalem]{ulem} 

\usepackage{mathtools}
\DeclarePairedDelimiter{\ceil}{\lceil}{\rceil}

\usepackage[binary-units=true]{siunitx} 
\usepackage{booktabs}
\usepackage{breqn}
\usepackage{algorithm}
\usepackage[noend]{algpseudocode}
\algnewcommand{\LeftComment}[1]{\Statex \(\triangleright\) #1} 
\algnewcommand\algorithmicinput{\textbf{Input}}
\algnewcommand\algorithmicoutput{\textbf{Output}}
\algnewcommand\Input{\item[\algorithmicinput]}%
\algnewcommand\Output{\item[\algorithmicoutput]}%
\algdef{SE}[VARIABLES]{Variables}{EndVariables}
   {\algorithmicvariables}
   {\algorithmicend\ \algorithmicvariables}
\algnewcommand{\algorithmicvariables}{\textbf{Global variables}}

\usepackage{xcolor}
\usepackage{ifthen}
\usepackage{tikz}

\usepackage{graphicx}
\usepackage{textcomp}
\usepackage{xcolor}
\def\BibTeX{{\rm B\kern-.05em{\sc i\kern-.025em b}\kern-.08em
    T\kern-.1667em\lower.7ex\hbox{E}\kern-.125emX}}
\begin{document}

\title{Generalized Data Placement Strategies for Racetrack Memories} 

\author{\IEEEauthorblockN{Asif Ali Khan, Andr{\'e}s Goens, Fazal Hameed and Jeronimo Castrillon}
\IEEEauthorblockA{Chair for Compiler Construction \\
Technische Universit\"at
Dreden, Germany \\
\{asif\_ali.khan, first.last\}@tu-dresden.de}
}

\maketitle

\begin{abstract}
Ultra-dense non-volatile \emph{racetrack memories} (RTMs) have been investigated at various levels in the memory hierarchy for improved performance and reduced energy consumption. 
However, the innate \emph{shift} operations in RTMs hinder their applicability to replace low-latency on-chip memories. 
Recent research has demonstrated that intelligent placement of memory objects in RTMs can significantly reduce the amount of shifts with no hardware overhead, albeit for specific system setups.
However, existing placement strategies may lead to sub-optimal performance when applied to different architectures.
In this paper we look at generalized data placement mechanisms that improve upon existing ones by taking into account the underlying memory architecture and the timing and liveliness information of memory objects.
We propose a novel heuristic and a formulation using genetic algorithms that optimize key performance parameters. 
We show that, on average, our generalized approach improves the number of shifts, performance and
energy consumption by 4.3$\times$, 46\% and 55\% respectively compared to the state-of-the-art. 
\end{abstract}


\section{Introduction}
\label{sec:intro}
The increasing capacity requirements along with the quest for higher performance and lower energy consumption have made memory system design extremely challenging. 
Traditional SRAM and DRAM technologies are unable to meet these antithetical requirements of today's applications due to larger cells and higher leakage power. 
On the contrary, emerging \emph{non-volatile memory} (NVM) technologies such as STT-RAM, phase change memory, magnetic RAM and \emph{racetrack memory} (RTM)~\cite{stuart4.0} offer a promising solution to fulfill these conflicting requirements.  
Recently, RTM has emerged as a leading contender due to its unprecedented capacity, energy efficiency and improved latency~\cite{Survey_Memories, stuart4.0}.  
For a feature size of $F$, the cell size of RTM is \~2$F^2$ whereas for STT-RAM and PCM cell sizes are \~6-50$F^2$ and \~4-12$F^2$ respectively.
Due to these promising characteristics, recent research advocate using RTM at various levels in the memory hierarchy~\cite{RTM_survey}.

\begin{figure}[tbh]
\centering
\includegraphics[scale=0.50]{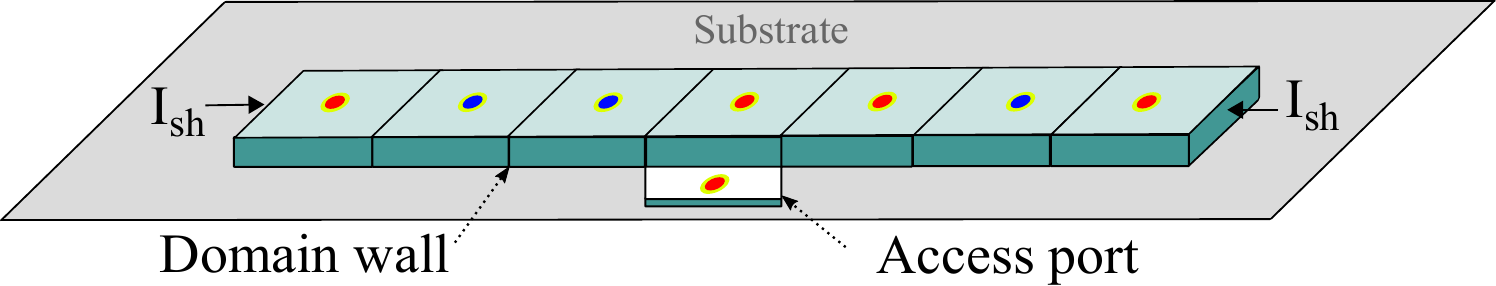}
\caption{RTM cell structure (red and blue dots on the nanowire represent upward and downward magnetization directions respectively)}
\label{fig:RTMOrg}
\end{figure}

A single RTM cell is a magnetic nanowire -- called \emph{nanotrack} -- that can store up to 100 domains where each domain represents a single bit~\cite{stuart4.0}. 
Each RTM nanotrack is equipped with one or more \emph{access ports} that perform read/write operations (cf. Fig.~\ref{fig:RTMOrg}).
To access a domain in a nanotrack, the relevant domain must be \emph{shifted} and aligned to the access port. 
Typically, multiple nanotracks are grouped together into \emph{Domain Block Clusters} (DBCs) to overlap the access transistor's footprint and thus effectively use the chip area budget. 
These \emph{shift operations} not only induce latency and energy overheads but also lead to variable access latencies, making RTM controller design especially challenging.

Recent literature suggests that intelligent placement of memory objects in a DBC substantially reduces the amount of RTM shifts (up to 50\%), improving both latency and energy consumption~\cite{Chen2016, shiftsreduce}. 
These initial solutions showed promising results for simplified system setups.
For instance, the heuristics in~\cite{Chen2016, shiftsreduce} provide data placement solution for a single DBC and the multi-DBC heuristic in~\cite{Chen2016} assumes a fixed multi-port architecture. 
In addition, the multi-DBC heuristic in~\cite{Chen2016} ignores valuable information of memory traces such as timing and liveliness information of memory objects, leading to sub-optimal solutions. 
In this paper, we propose generalized data placement strategies that are independent of the RTM architecture and exploit the timing information in memory traces before deciding the layouts. The proposed solutions carefully distribute memory objects across DBCs and judiciously assign exact locations to objects within DBCs.
Concretely, we make the following contributions:

\begin{enumerate}
  \item A novel fast heuristic that analyzes the memory trace for objects with \emph{disjoint} lifespans and steers disjoint and non-disjoint memory objects to separate DBCs. This separation significantly improves temporal locality of the memory objects and reduces the number of shifts.
  \item A more time-consuming heuristic based on genetic algorithms that achieves near optimal results.
  \item A thorough analysis of the interplay of different solutions for inter- and intra-DBC placements of  memory objects. We also analyze the impact of increasing the number of DBCs on performance, energy and area.
\end{enumerate}

\section{Background}
\label{sec:background}
This section presents a detailed description of RTM architectures and their organization. 
It also provides background on both inter and intra-DBC data placements and highlights the importance of inter-DBC memory objects distribution. 

\begin{figure}[tbh]
\centering
\includegraphics[scale=0.50]{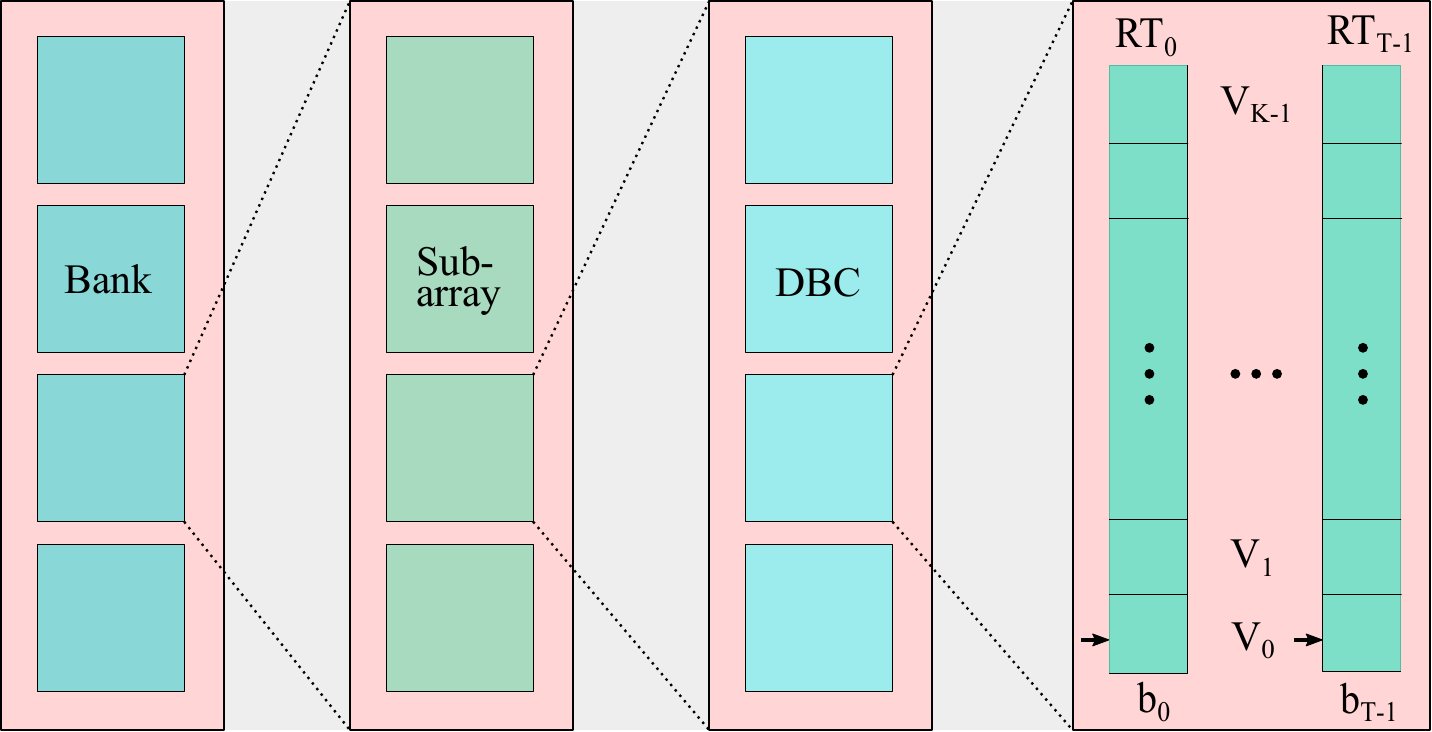}
\caption{RTM architecture}
\label{fig:DBCOrg}
\end{figure}

\subsection{RTM architecture}
\label{ss:RtmArch}
Fig.~\ref{fig:DBCOrg} illustrates a common RTM architecture. 
Similar to other memory technologies, RTM consists of multiple banks where each bank contains one or more subarrays. 
Each subarray in RTM comprises multiple DBCs, each of them with $T$ nanotracks. 
A nanotrack stores $K$ domains (i.e., bits) and has one or more access ports to perform read/write operations (cf. Fig.~\ref{fig:RTMOrg}).
Typically, data is stored in a bit-interleaved fashion so that all $T$ bits of a memory object are kept in the $T$ nanotracks of a DBC as illustrated in Fig.~\ref{fig:DBCOrg}.
To access a memory object, bits are shifted in a lock-step fashion until they are aligned to the access port positions~\cite{tapcache}.

\subsection{State-of-the-art data placement in RTMs}
\label{ss:DataPlace}
The data placement problem in RTMs can be artificially decomposed into two subproblems, 
the inter-DBC data placement problem which steers program variables (or memory objects) to different DBCs, and the intra-DBC 
data placement problem.
The latter finds a suitable placement of the variables in a particular DBC.
To this end, heuristics are employed which aim to find near-optimal intra-DBC data placement in reasonable time~\cite{shiftsreduce, Chen2016}.
These solutions pay little to no attention to the inter-DBC distribution of memory objects.

Intra-DBC placement heuristics are inspired by the well-known heuristics for single offset assignment~\cite{OffsetStone,tsp_soa}.
The problem consists in assigning a set of variables $V = \{v_1,\dots,v_n\}$ a location in the memory, based on an access trace referred to as \emph{access sequence} $S = (s_1,\dots,s_{k})$ where $s_1,\ldots,s_{k} \in V$. 
The access sequence is typically summarized in a weighted undirected \emph{access graph}.
Vertices in the access graph represent variables while an 
edge $ e = \{u, v\}$ expresses that the variables corresponding to $u$ and $v$ were consecutively accessed in $S$.
The edge weight $w_{uv}$ models the number of such consecutive accesses. 
Finally, the access frequency of a variable $u$ is the number of times $u$ is accessed in $S$.

Based on the information in the access graph, heuristics aim to place memory objects within a DBC to maximize the likelihood that consecutive accesses in $S$ access the same or nearby locations in a DBC, resulting in a reduced shift cost.
The shift cost between two accesses $u$ and $v$ in $S$ is the absolute difference of their exact locations in a DBC, as this corresponds to the number of shifts an RTM controller will need to execute in order to access $u$ after accessing $v$~\cite{shiftsreduce, Chen2016}.

%

State-of-the-art heuristics mainly focus on addressing the intra-DBC data placement problem. What has got little attention is the inter-DBC distribution of memory objects which is equally important because, as evidenced by Sec.~\ref{ss:RtmArch},  typical RTM organizations have more than one DBCs. 
Chen et al.~\cite{Chen2016} briefly explain the inter-DBC data placement and present a heuristic that distributes memory objects across DBCs based on their access frequencies (cf. Sec.~\ref{subsec:chen-AFD}). 
However, we argue that access frequencies alone are not sufficient to find a good memory layout. 
Memory objects with disjoint lifespans when placed in the same DBC while maintaining their access order substantially reduces the amount of shifts. 
Similarly, Chen's multi-DBC heuristic is designed for RTMs with two or more access ports per track. 
The next section discusses generalized data placement solutions that are independent of the number of ports and use timing and liveliness information of the memory objects to find efficient inter- and intra-DBC placement. 

\section{Generalized data placement in RTM} 
\label{sec:proposed}
This section describes our proposed solutions for data placement in RTMs
after explaining the state-of-the-art inter-DBC placement technique.

\subsection{Baseline inter-DBC placement}
\label{subsec:chen-AFD}
To the best of our knowledge, the current best inter-DBC data placement heuristic was proposed  in~\cite{Chen2016}.
The \emph{Access Frequency based Distribution} (AFD) heuristic initially sorts the variables in $V$ 
in descending order of their access frequencies.
It then iteratively selects variables and distributes them to DBCs in a round-robin manner.
The basic idea is to place frequently accessed variables as 
close as possible to reduce the shift overhead.

Fig.~\ref{fig:Example} shows a placement example to two DBCs for the variables in 
Fig.~\ref{fig:Example}-(a) and access sequence in Fig.~\ref{fig:Example}-(b).
The AFD heuristic, in Fig.~\ref{fig:Example}-(c), assigns variables $a$, $g$, $b$, $d$, and $h$ to
$DBC_0$ and $e$, $i$, $c$, and $f$ to $DBC_1$.
Since accesses are partitioned between $DBC_0$ and $DBC_1$, the 
access sequence $S$ is split into two disjoint subsequences $S_0$ and $S_1$.
Applying the AFD heuristic to the sample access sequence incurs 24 and 15 shifts for accessing variables
in $S_0$ and $S_1$ respectively.
As a result, the overall shift cost amounts to 39.
In the next section, we show that by better exploiting the access order and timing information an improved placement can be obtained.

\subsection{Sequence-aware inter-DBC distribution}
\label{subsec:proposed-heu}

\begin{figure}[tbh]
\centering
\includegraphics[scale=0.59]{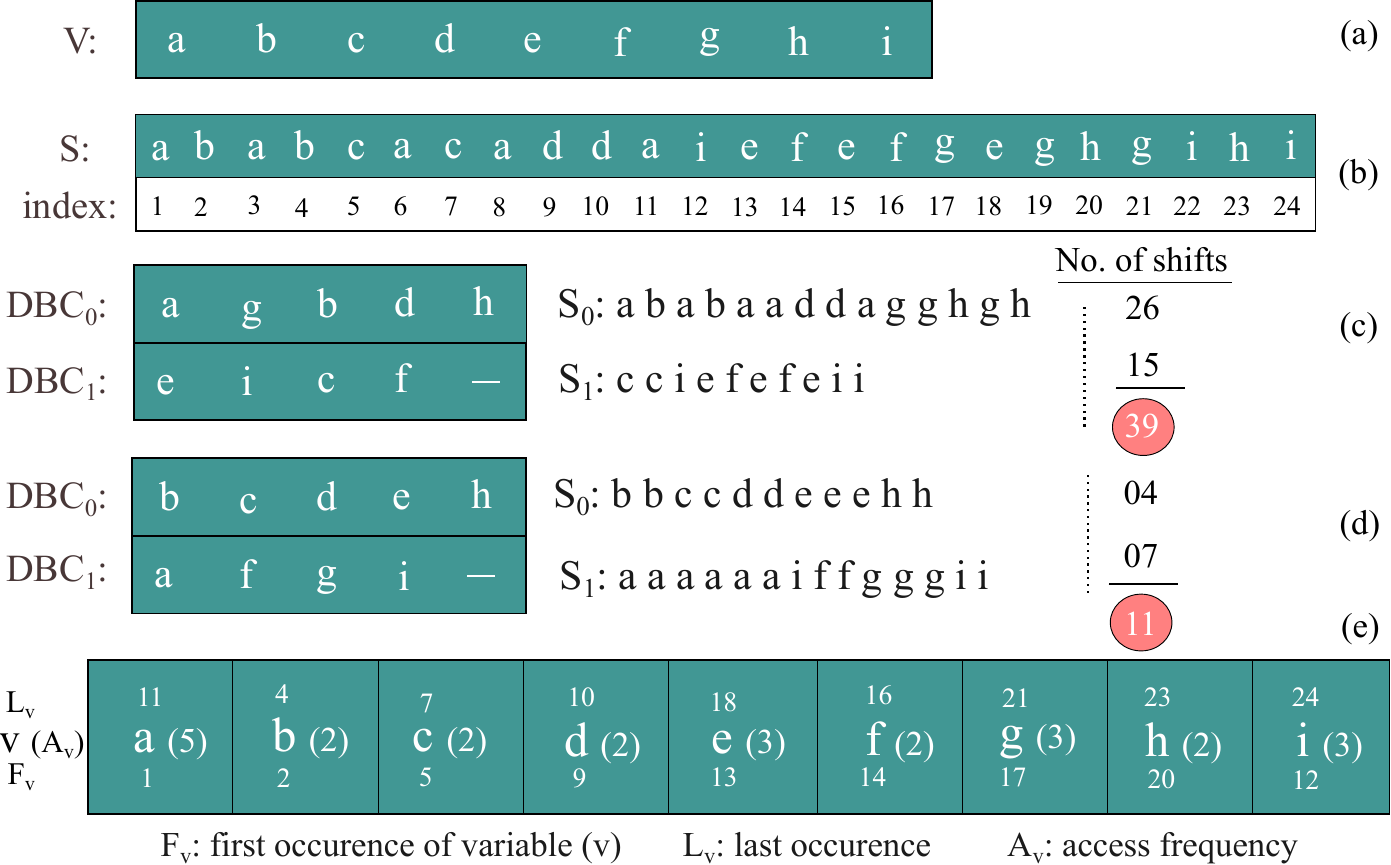}
\caption{Example showing (a) Variable set (b) Access sequence and the time of occurrence of each access (c) AFD placement~\cite{Chen2016} (d) Sequence-aware placement (e) Timing and access frequency of each variable}
\label{fig:Example}
\end{figure}

The access graph, commonly used to summarize the access sequence, discards the order and the timing information of memory objects. Our heuristic takes these information into account and combines them with the access frequencies for a more efficient inter-DBC placement.
Two variables $u$ and $v$ are said to have \emph{disjoint} lifespans if the last occurrence of $u$ in $S$ is before the first occurrence of $v$ in $S$ and vice versa.
The \emph{lifespan} of a variable is then defined as the absolute difference of its first and last occurrences.
For instance, in the access sequence $S$ in Fig.~\ref{fig:Example}-(b), the lifespan of variable $b$ is 2 ($4-2$) and variables $b$ and $c$ have disjoint lifespans.

Our heuristic exploits the fact that $l$ disjoint variables, if stored in the same DBC while respecting their access order, require at most $l - 1$ RTM shifts. 
This implies that once an access port is aligned to one of the $l$ variables in the DBC, the following accesses to the same variable will not incur any shifts at all.
Accessing the next variable in the same DBC will always incur only a single shift. 
To explain this further, let us consider the sample access sequence from Fig.~\ref{fig:Example}-(b) and the corresponding access frequency and timing information from Fig.~\ref{fig:Example}-(e).
Our heuristic extracts a list of disjoint variables having  maximum sum of access frequencies by performing the liveliness analysis on all memory objects. 
In other words, the heuristic picks a variable combination that maximizes the number of self accesses which in turn reduces the total amount of shift operations. 
For the illustrating example, our heuristic analyses memory object $a$ by comparing its access frequency with the sum of access frequencies of all those objects that lie in the lifespan of $a$ ($b,c,d$). If the access frequency of $a$ (5) is greater than the sum of access frequencies of all those memory objects (6), the heuristic appends $a$ to the list of disjoint variables otherwise it moves to the next object and repeats this exact same process. For the illustrating example, our heuristic selects combination $b$, $c$, $d$, $e$, $h$ having sum of access frequencies equal to 11.

Variables in the selected combination are allocated to the same DBC (i.e., $DBC_0$ in the illustrating example) in their access order.
Note however that this preservation of access order is only restricted to the DBC that stores variables with disjoint lifespans.
For other DBCs, heuristics such as~\cite{Chen2016, shiftsreduce} are employed to find an efficient intra-DBC placement. 
The leftover variables (i.e., $a$, $f$, $g$, and $i$) are assigned to the remaining DBCs (i.e., $DBC_1$), which is shown in Fig.~\ref{fig:Example}-(d).
Compared with the AFD solution~\cite{Chen2016} in Fig.~\ref{fig:Example}-(c), the shift cost is reduced from 39 to 11 (i.e., 3.54$\times$ shifts improvement).

\begin{algorithm}[tbh]
\caption{Proposed data distribution heuristic}
\begin{algorithmic}[1]
\Input: Access sequence $S$, list of variables $V$ and $q$ DBCs each having $N$ empty locations 
\Output: Final data distribution across all DBCs
\State \Comment{Initialize access freq., first and last accesses} 
\ForAll {$v \in V$} $A_v = \sum_{u\in S, u=v}1$ \EndFor
\ForAll {$v \in V$} $F_v = \min{\{i\in\{1,\dots,|S|\} \mid S_i = v\}}$ \EndFor
\ForAll {$v \in V$} $L_v = \max{\{i\in\{1,\dots,|S|\} \mid S_i = v\}}$ \EndFor
\State $V_{ndj} \gets $Variables $V$ sorted in the ascending order of $F_v$  \label{SEM:sort}
\State $V_{dj} \gets \emptyset$  \label{SEM:EmptySetInitialize}
\State $ t_{min} \gets 0$    
  \ForAll {$v \in V_{ndj}$}  \label{SEM:DisjointStart}
    \If {$F_{v} > t_{min}$} 
      \If {$A_v >  \sum_{u \in V_{ndj}:F_u > F_v, L_u < L_v}$}
            \State $V_{dj} \gets V_{dj} \cup \{ v \}$ \label{SEM:Appendvi}
            \State $V_{ndj} \gets V_{ndj} \setminus \{ v \}, t_{min} \gets L_v$ 
      \EndIf
    \EndIf
  \EndFor \label{SEM:DisjointEnd}

\State $ K \gets \ceil*{\frac{\lvert V_{dj} \rvert}{N}} $ \label{SEM:disjointDBCs} 
  

    \While{ $\lvert V_{dj} \rvert > 0$} \label{SEM:AssignToDBCsStart}
    \For{ $i \gets 1, \ldots, K$}
       \State $ v^* \gets \text{argmin}_{v\in V_{dj}}F_v$
       \State $DBC_{i}.\operatorname{append}(v^*), V_{dj} \gets V_{dj} \setminus \{v^*\}$ 
       \EndFor 
    \EndWhile

    \While{ $\lvert V_{ndj} \rvert > 0$} 
    \For{ $i \gets K+1, \ldots, q$}
       \State $ v^* \gets \text{argmax}_{v\in V_{ndj}}A_v$
       \State $DBC_{i}.\operatorname{append}(v^*), V_{ndj} \gets V_{ndj} \setminus \{v^*\}$ 
    \EndFor
    \EndWhile \label{SEM:AssignToDBCsEnd}

\For{ $i \gets K+1, q$} \label{SEM:PostOptStart}
\State Apply Chen~\cite{Chen2016} or ShiftsReduce~\cite{shiftsreduce} on DBC$_i$ 
\EndFor \label{SEM:PostOptEnd}
%

\end{algorithmic}
\label{algo:SEM}
\end{algorithm}

Algorithm~\ref{algo:SEM} shows the pseudocode of our proposed data placement heuristic. 
The heuristic maintains two sets of variables, $V_{dj}$ and $V_{ndj}$ storing disjoint and non-disjoint variables respectively. 
Similarly, the variables $A_v, F_v$ and $L_v$ store the access frequency, first and last occurrence information of all variables in $V$ respectively. 
Initially, $V_{ndj}$ stores all variables in $V$ (line~\ref{SEM:sort}), and when we iterate through it (line~\ref{SEM:DisjointStart}) we do so in the ascending order of their first occurrences $F$. 
$V_{dj}$ is initialized as an empty set (line~\ref{SEM:EmptySetInitialize}).
The algorithm then iteratively selects variables $v_i$ from $V_{ndj}$, examines disjointness and appends only those variables to $V_{dj}$ that maximize the number of self accesses (lines~\ref{SEM:DisjointStart}-\ref{SEM:DisjointEnd}). 
The variable $K$ (line~\ref{SEM:disjointDBCs}) computes the number of DBCs required for storing disjoint variables ($V_{dj}$). 
The variables in $V_{dj}$ are assigned to DBCs $1 \to K$ and $V_{ndj}$ to the remaining $(q-K)$ DBCs (lines~\ref{SEM:AssignToDBCsStart}-\ref{SEM:AssignToDBCsEnd})
where $q$ represents the total number of DBCs. 
Finally, lines~\ref{SEM:PostOptStart}-\ref{SEM:PostOptEnd} apply the single DBC heuristics from~\cite{Chen2016, shiftsreduce} to optimize within DBC placement of program variables.

\subsection{Genetic algorithms for data placement in RTM}
\label{subsec:ga-rs}
Practicality in compilers demands fast-executing heuristics, like the one we propose.
However, as a baseline to evaluate heuristics it is extremely useful to know the optimal solution to a problem.
Given that finding an optimal multi-DBC placement is an NP complete problem~\cite{Chen2016}, 
we present a formulation using genetic algorithms (GAs) for finding near-optimal results that 
serves as baseline.

In our formulation, individuals represent the final variable placements (both inter- and intra-DBC).
We represent them as lists of DBC assignments $I = (DBC_1,\ldots,DBC_q)$, whereby each DBC assignment $DBC_i = (v^{(i)}_1,\ldots,v^{(i)}_{|DBC_1|})$ is in turn a list with the variable placements in the selected order.
The fitness value of an individual is the shifts cost of that variable placement.
Our GA formulation uses a $\mu + \lambda$ algorithm, whereby we produce $\lambda = 100$ offspring each iteration and select $\mu = 100$ individuals for the next generation.
The individual selection follows a tournament model, selecting the individual with the best fitness value out of $4$ randomly-selected individuals in the population. 
These parameters were chosen to get best-effort results in a reasonable time in our implementation.

To produce offspring, we use a $2$-fold crossover on the individuals. Let $I,J$ be two individuals. Let $V = {v_1, \ldots, v_n}$ where the $v_i$ are indexed in the same order as they appear in the sequence $S$.
We randomly select two variables $v_{f}, v_{l}, f < l, \in V$ as crossover points, and separate $V$ into the disjoint union $V = V_{\text{swap}} \cup V_{\text{leave}}$, where $V_{\text{swap}} = \{v_f,v_{f+1},\ldots,v_l\}$
and $V_{\text{leave}} = V \setminus V_{\text{swap}}$. Then we swap the assignments of variables in $V_{\text{swap}}$ between $I$ and $J$:
\begin{align*}
  \forall v \in V_{\text{swap}}, \text{ s.t. } v \in DBC^I_r, v \in DBC^J_s \text{ and } r \neq s: \\
  DBC^I_r.\operatorname{remove}(v), DBC^I_s.\operatorname{append}(v) \\
  DBC^I_s.\operatorname{remove}(v), DBC^I_r.\operatorname{append}(v), 
\end{align*}
This ensures that the within-DBC variable placements that are not swapped are kept and that both new individuals are still valid placements.
A mutation, on the other hand, selects one of three possible mutations at random:
\begin{itemize}
  \item Move a variable from one DBC to another, placing it at the end of the new DBC and leaving the rest of the variables in the same order.
  \item Transpose two variables in a single DBC.
  \item Apply random permutation to each DBC.
\end{itemize}

The first mutation slightly modifies the inter-DBC placement. The second and third mutations change the permutation within a single DBC.
Since the third option is more destructive, we skew the probability so that it is less likely to happen with in a ratio of $10:3$.
These mutations make sure that both, the mutated assignments are still correct assignments, and for any two possible assignments, there is a series of mutations taking one to the other.
This way we can explore the whole design space of assignments.
For comparison, we also implemented a random-walk search which generates random placement of variables to DBCs and the create random permutations within every DBC, selecting the best individual.


\section{Evaluation}
\label{sec:methodology}
This section describes the experimental setup and compares our proposed solutions to the state-of-the-art.

\subsection{Experimental setup}
\label{subsec:setup}
For evaluation, we use the open-source RTSim simulator~\cite{rtsim} that takes application memory traces and produces latency and energy results. 
We simulate all 30 benchmarks of the OffsetStone benchmark suite~\cite{OffsetStone}, including real-world application domains such as image, signal and video processing, and control-dominated applications such as GZIP, BISON, Flex and CPP.
Benchmarks vary in terms of number of access sequences, number of program variables per sequence (i.e., 1 to 1336) and the length of access sequences (1 to 3640).

The latency, energy and area numbers for different RTM configurations are obtained from the  destiny circuit simulator~\cite{destiny} and are listed in Table~\ref{tab:params}. 
These values also include the latency incurred and the energy consumed by the DBC/domain decoders, access ports, multiplexers, write and shift drivers.
All iso-capacity RTM configurations are chosen so that each of them has different number of DBCs (i.e, 2 to 16) 
and domains per DBC (i.e., 64 to 512).

\begin{table}[t]
\centering
\caption{Memory system parameters (\SI{4}{\kibi\byte} RTM, \SI{32}{\nano\meter}, $32$ tracks $/$ DBC)}
\centering
\label{tab:params}
  \begin{tabular}{c|c|c|c|c}
\toprule
Number of DBCs & 2 & 4 & 8 & 16\\
Number of domains in a DBC & 512 & 256 & 128 & 64\\
Leakage power [\SI{}{\milli\watt}] & 3.39 & 4.33 & 6.56 & 8.94\\
Write energy [\SI{}{\pico\joule}] & 3.42 & 3.65 & 3.79 & 3.94\\
Read energy [\SI{}{\pico\joule}] & 2.26 & 2.39 & 2.47 & 2.54\\
Shift energy [\SI{}{\pico\joule}] & 2.18 & 2.03 & 1.97 & 1.86\\
Read latency [\SI{}{\nano\second}] & 0.81 & 0.84 & 0.86 & 0.89\\
Write latency [\SI{}{\nano\second}] & 1.08  & 1.14 & 1.17 & 1.20\\
Shift latency [\SI{}{\nano\second}] & 0.99 & 0.92 & 0.86 & 0.78\\
Area [\SI{}{\mm}$^{\text{2}}$] & 0.0159 & 0.0186 & 0.0226 & 0.0279 \\
\bottomrule
\end{tabular}
\end{table}

We evaluate six different data placement solutions as listed below. 
Unless otherwise stated, all results are normalized to the results of the genetic algorithm.

\begin{itemize}
\item \emph{AFD-OFU:} The baseline inter-DBC data placement heuristic~\cite{Chen2016}. The intra-DBC placement of variables is based on their \emph{order of first use} (OFU).
\item \emph{DMA-OFU:} Our proposed heuristic separating \emph{disjoint memory accesses} (DMA) from non-disjoint accesses (cf. Sec.~\ref{subsec:proposed-heu}) with OFU assignment. 
\item \emph{DMA-Chen:} Our proposed heuristic paired with the intra-DBC optimization heuristic (Chen~\cite{Chen2016}, single DBC).
\item \emph{DMA-SR:} Our proposed heuristic paired with the ShiftsReduce heuristic~\cite{shiftsreduce}.
\item \emph{GA:} Our proposed genetic algorithm (cf. Sec.~\ref{subsec:ga-rs}).
\item \emph{RW:} A random walk search (cf. Sec.~\ref{subsec:ga-rs}). 
\end{itemize}

We execute \emph{GA} for $200$ generations, and \emph{RW} for $60000$ iterations, which is the upper bound on the number of individuals that could be evaluated by \emph{GA} with these parameters.

\subsection{Analysis of heuristics: Reduction in shifts}
\label{res:shifts}
\begin{figure*}[tbh]
\centering
\resizebox{0.92\textwidth}{!}{\input{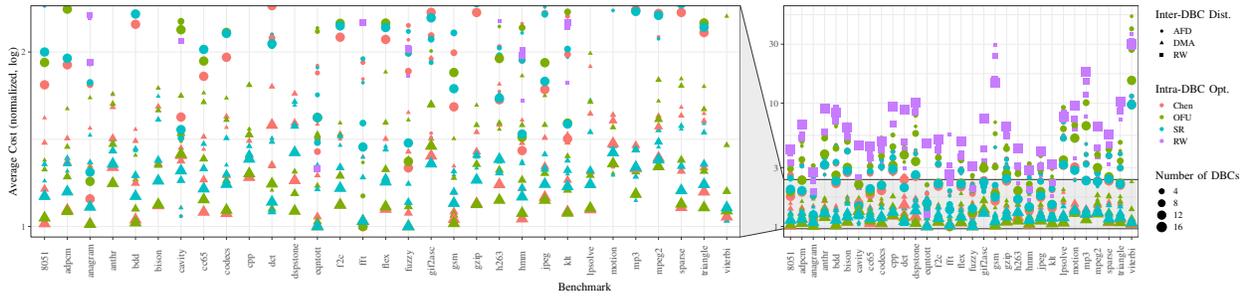}}
\caption{Shifts improvement in our proposed solutions for various RTM configurations}
\label{fig:res:shifts}
\end{figure*}

Fig.~\ref{fig:res:shifts} shows the normalized shift improvement of our proposed solutions compared to the baseline. 
The results are normalized to the costs obtained from the placement in GA (i.e. the costs for GA are always $1)$. 

As can be seen, our proposed heuristic significantly reduces the number of RTM shifts.
More concretely, the reduction as expressed by the \emph{geometric mean} over all benchmarks is 2.4$\times$, 2.9$\times$, 2.8$\times$ and 1.7$\times$ compared to AFD-OFU for 2, 4, 8, and 16 DBC RTM configurations respectively.
DMA-Chen and DMA-SR further diminish the amount of shifts by (1.8$\times$, 1.6$\times$, 1.3$\times$, 1.4$\times$) and (2.0$\times$, 1.8$\times$, 1.5$\times$, 1.6$\times$) for (2, 4, 8, 16) DBCs respectively. 
Fig.~\ref{fig:res:shifts} also demonstrates that the shift reduction is less pronounced when more DBCs are employed.
This is because an increase in the number of DBCs leads to a more sparse variable distribution, making the shift problem less severe. 
For the same reason, the gain from intra-DBC placement is less prominent as we increase the DBC count.

\emph{RW} results serve to put the \emph{GA} results in perspective, as \emph{RW} evaluated more individuals for every benchmark. 
To asses how far the heuristics are from the optimal solution, we executed \emph{GA} significantly longer for the benchmark with the largest access sequence.
After $2000$ generations, the result from the best variant of the heuristics was around $38\%$ worse than the best solution found by the \emph{GA}. 
This indicates that our solutions are likely within a reasonable range of the optimum, less than an order of magnitude.

The simulation results also suggest that our distribution heuristic consistently performs well irrespective of the DBC count and the intra-DBC optimization. 
In fact, it provides a promising base for the Chen and ShiftsReduce heuristics to further improve its performance and minimize the shift cost. 
For the above reasons, we expect our heuristic to perform well with future optimization policies as well.

\subsection{Overall performance and energy analysis}
\label{res:perf}

We also compare the heuristics in terms of latency and energy consumption.
DMA-OFU improves the RTM access latency by 50.3\%, 50.5\%, 33.1\% and 10.4\% for 2, 4, 8 and 16 DBC configurations respectively. 
DMA-Chen and DMA-SR further improve the latency by (68.1\%, 60.1\%, 36.5\%, 13.4\%) and (70.1\%, 62\%, 37.7\%, 14.6\%) for (2, 4, 8, 16) DBCs respectively. 
The latency gain primarily stems from reduced number of RTM shifts which reduces the RTM access latency and ultimately the overall runtime. 

\begin{figure}[tbh]
\centering
\includegraphics[scale=0.450]{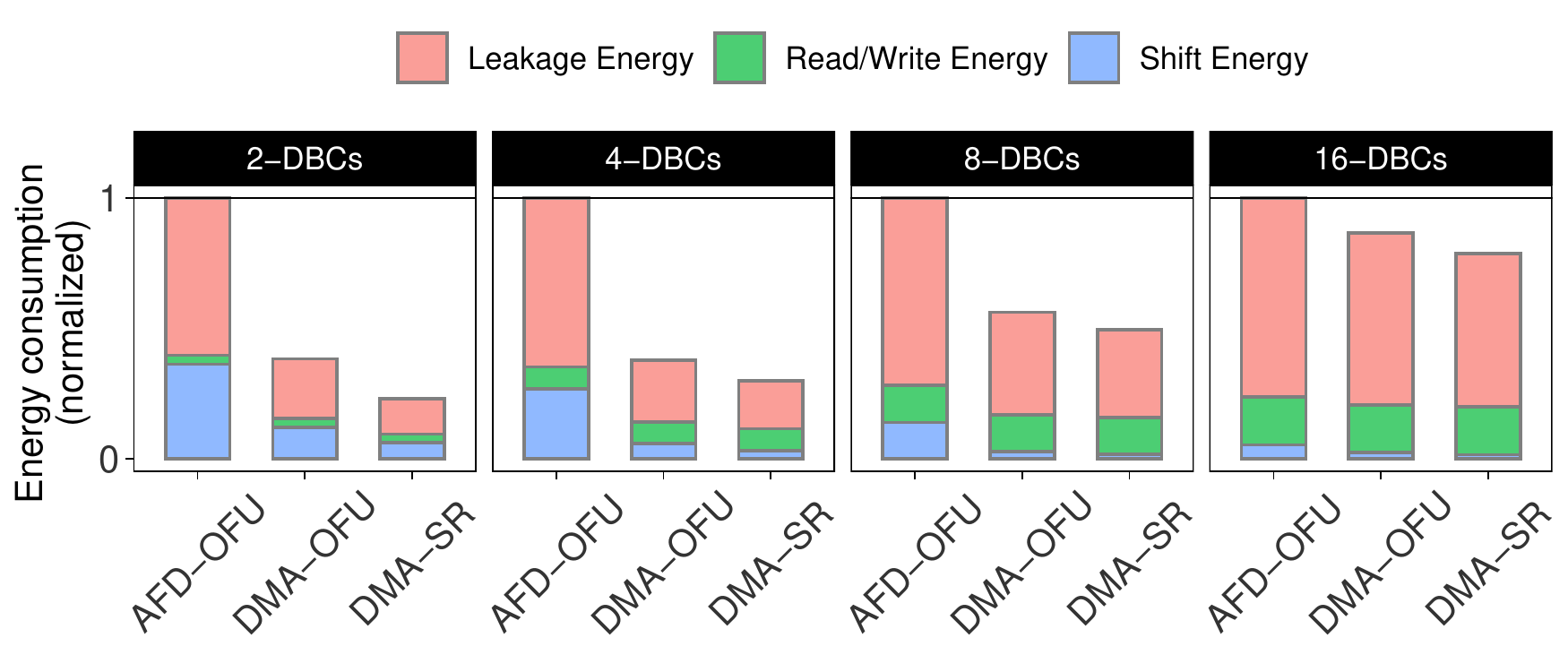}
\caption{Overall energy consumption normalized to the baseline AFD-OFU}
\label{fig:res:energy}
\end{figure}

Fig.~\ref{fig:res:energy} highlights the significant reduction in the total energy consumed by DMA-OFU (61\%, 62\%, 44\%, 13\%) and DMA-SR (77\%, 70\%, 50\%, 21\%) relative to AFD-OFU for (2, 4, 8, 16) DBCs respectively. 
By breaking down the energy consumption into leakage energy, read/write and shift energy, we observe that 
(1) the gain in shift energy is proportional to the reduction in the number of shifts,  
(2) leakage energy becomes more significant as the number of DBCs increases (cf. Table~\ref{tab:params}), and 
(3) in both DMA-OFU and DMA-SR, the leakage energy marks a substantially drop-down.
Our analysis suggest that the latter is due to the runtime reduction.
The performance and energy results indicate that our distribution heuristic greatly outperforms AFD distribution in both metrics.

\begin{figure}[tbh]
\centering
\includegraphics[scale=0.450]{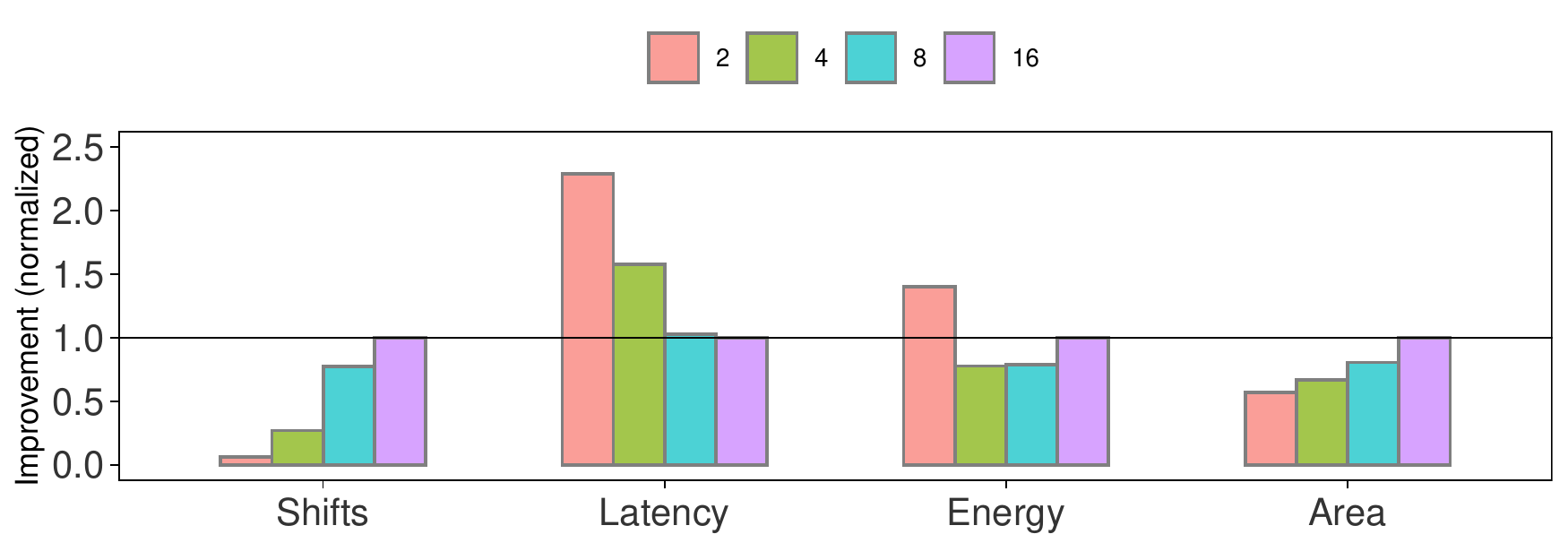}
\caption{Impact of varying the number of DBCs for DMA-SR configuration}
\label{fig:res:overall}
\end{figure}

Fig.~\ref{fig:res:overall} shows the trade-off among various parameters for the best performing DMA-SR configuration as we increase the number of DBCs from 2 to 16.
The area values indicate a clear rising trend with the increase in the number of DBCs (or ports).
The major reason is that, access ports have a larger footprint compared to other components of an RTM.
In terms of energy consumption, Fig.~\ref{fig:res:overall} demonstrates that a 2-DBC RTM is not competitive 
due to its high shift energy contribution (Fig.~\ref{fig:res:energy}).
In this case, the positive impact of a reduced leakage power is negatively offset by increase in the 
shift energy.
We also notice that the latency and the shift improvement diminish significantly with an increased DBC count.
As a consequence, the shift energy contribution becomes less prominent and in turn 
a 16-DBC RTM consumes more energy than a 4-DBC or 8-DBC variant.

\section{Related work}
\label{sec:related_work}
RTMs have been employed at various levels in the memory hierarchy to demonstrate 
its performance and energy benefits.
For instance, it has been shown that shifts reduction to the bare minimum in
RTM scratchpad improves the performance and the energy saving by 24\% and 74\% 
respectively compared to an iso-capacity SRAM for tensor contraction~\cite{Khan_LCTES}.
Likewise, similar benefits have been demonstrated at higher levels, e.g., caches~\cite{ArrayOrgDWM2016,tapcache} and main memory~\cite{RTMMemDSE2016}.

Many techniques have been proposed in the past to mitigate the negative impact of RTM shift overhead. 
These include data compression~\cite{xu2015}, reconfigurability of RTM in terms of deactivating (or activating) rarely (or highly) used domains, runtime data swapping~\cite{sun2013}, proactively aligning the likely accessed domains to the port positions~\cite{predictor_based_preshifting, mao2015, tapcache, sun2013}, and intelligent instruction~\cite{GOGAL_SHRIMP} and data placement~\cite{Chen2016, shiftsreduce, Khan_LCTES, rm_zpu, rm_gpu_rf}.
Among these proposals, data placement has demonstrated significant benefits with trivial or no overheads.
These techniques primarily focus on intra-DBC variable assignment to curtail the shift overhead~\cite{shiftsreduce, Chen2016}.
We showed that the most recent inter-DBC placement from~\cite{Chen2016} leads to sub-optimal performance as it only considers the access frequency of individual variables but ignores the variable liveliness information (cf. Sec.~\ref{ss:DataPlace},~\ref{subsec:chen-AFD}).

Hardware- and software-guided data placement techniques have also been used in the past in the context of other NVMs and hybrid memory systems~\cite{rthms, UBHM_2017} to hide higher NVM write latency.
However, for RTMs we aim at finding a layout for memory objects that minimizes the number of RTM shifts, a problem that does not pertain to other random access volatile or non-volatile memories.
As a result, these data placement solutions are not applicable to RTMs.

\section{Conclusions and outlook}
\label{sec:conclusion}
In this paper, we presented a novel solution for generalized data placement in RTM. 
We proposed a novel heuristic that analyzes the lifespans of memory objects and steers them to DBCs with the objective to minimize the total number of shifts.
Our evaluation showed a substantial reduction of shifts by 4.3$\times$ compared to the state of the art heuristic. 
The average improvements in latency and energy consumption across all benchmarks and all configurations was  of 46\% and 55\% respectively.
We demonstrated that our heuristic consistently outperformed the state-of-the-art for different number of DBCs and can be paired with existing single DBC data placement solutions.
Our formulation as genetic algorithms, with customized genetic operators and our heuristic result as initial population, showed that the heuristic results, in terms of the number of shifts, are likely within an order of magnitude of the optimum. 
In future work, we plan to explore placement of more than one sets of disjoint variables in the same DBC and in different DBCs and their integration with non-disjoint variables in a way that further reduces the overall shift cost.

\section*{Acknowledgments}
This work was partially funded by the German Research Council (DFG) through the TraceSymm project CA 1602/4-1 and the Cluster of Excellence `Center for Advancing Electronics Dresden' (cfaed).


\bibliographystyle{abbrv}
\bibliography{Date20}

\begin{thebibliography}{10}

\bibitem{predictor_based_preshifting}
E.~Atoofian.
\newblock {Reducing Shift Penalty in Domain Wall Memory Through Register
  Locality}.
\newblock In {\em Proc. of the 2015 Int. Conf. on Compilers, Architecture and
  Synthesis for Embedded Systems}, pages 177--186, 2015.

\bibitem{Chen2016}
X.~Chen~et al.
\newblock {Efficient Data Placement for Improving Data Access Performance on
  Domain-Wall Memory}.
\newblock {\em IEEE Trans. Very Large Scale Integr. Syst.}, 24(10):3094--3104,
  Oct. 2016.

\bibitem{RTMMemDSE2016}
Q.~{Hu}, G.~{Sun}, J.~{Shu}, and C.~{Zhang}.
\newblock {Exploring Main Memory Design Based on Racetrack Memory Technology}.
\newblock In {\em 2016 International Great Lakes Symposium on VLSI (GLSVLSI)},
  pages 397--402, May 2016.

\bibitem{tsp_soa}
M.~J\"{u}nger and S.~Mallach.
\newblock Solving the simple offset assignment problem as a traveling salesman.
\newblock In {\em Proc. of the 16th Int. Workshop on Software and Compilers for
  Embedded Systems}, pages 31--39, 2013.

\bibitem{Khan_LCTES}
A.~A. Khan, N.~A. Rink, F.~Hameed, and J.~Castrillon.
\newblock {Optimizing Tensor Contractions for Embedded Devices with Racetrack
  Memory Scratch-pads}.
\newblock In {\em Proc. of the 20th Int. Conf. on Languages, Compilers, and
  Tools for Embedded Systems}, LCTES 2019, pages 5--18, 2019.

\bibitem{rtsim}
A.~A. {Khan et al.}
\newblock {RTSim: A Cycle-Accurate Simulator for Racetrack Memories}.
\newblock {\em IEEE Computer Architecture Letters}, 18(1):43--46, Jan 2019.

\bibitem{shiftsreduce}
A.~A. {Khan et al.}
\newblock {ShiftsReduce: Minimizing Shifts in Racetrack Memory 4.0}.
\newblock {\em arXiv e-prints}, Mar 2019.

\bibitem{rm_zpu}
H.~A. Khouzani~et al.
\newblock {A DWM-Based Stack Architecture Implementation for Energy Harvesting
  Systems}.
\newblock {\em ACM Trans. Embed. Comput. Syst.}, 16(5s):155:1--155:18, Sept.
  2017.

\bibitem{OffsetStone}
R.~Leupers.
\newblock {Offset Assignment Showdown: Evaluation of DSP Address Code
  Optimization Algorithms}.
\newblock In {\em Proceedings of the 12th International Conference on Compiler
  Construction}, CC'03, pages 290--302, 2003.

\bibitem{UBHM_2017}
Y.~{Li}, S.~{Ghose}, J.~{Choi}, J.~{Sun}, H.~{Wang}, and O.~{Mutlu}.
\newblock {Utility-Based Hybrid Memory Management}.
\newblock In {\em 2017 IEEE International Conference on Cluster Computing
  (CLUSTER)}, pages 152--165, September 2017.

\bibitem{rm_gpu_rf}
Y.~Liang and S.~Wang.
\newblock Performance-centric optimization for racetrack memory based register
  file on gpus.
\newblock {\em Journal of Computer Science and Technology}, 31(1):36--49, Jan
  2016.

\bibitem{mao2015}
H.~Mao~et al.
\newblock {Exploring Data Placement in Racetrack Memory Based Scratchpad
  Memory}.
\newblock In {\em 2015 IEEE Non-Volatile Memory System and Applications
  Symposium (NVMSA)}, pages 1--5, Aug 2015.

\bibitem{RTM_survey}
S.~Mittal.
\newblock A survey of techniques for architecting processor components using
  domain-wall memory.
\newblock {\em J. Emerg. Technol. Comput. Syst.}, 13(2):29:1--29:25, Nov. 2016.

\bibitem{Survey_Memories}
S.~Mittal, J.~S. Vetter, and D.~Li.
\newblock {A Survey Of Architectural Approaches for Managing Embedded DRAM and
  Non-Volatile On-Chip Caches}.
\newblock {\em IEEE Tran. on Parallel and Dist. Systems}, 26(6):1524--1537,
  June 2015.

\bibitem{destiny}
S.~Mittal, R.~Wang, and J.~Vetter.
\newblock {DESTINY: A Comprehensive Tool with 3D and Multi-Level Cell Memory
  Modeling Capability}.
\newblock {\em Journal of Low Power Electronics and Applications}, 7(3), 2017.

\bibitem{GOGAL_SHRIMP}
J.~Multanen~et al.
\newblock {SHRIMP: Efficient Instruction Delivery with Domain Wall Memory}.
\newblock In {\em Proc. of the Int. Symposium on Low Power Electronics and
  Design}, ISLPED '19, July 2019.

\bibitem{stuart4.0}
S.~Parkin and S.-H. Yang.
\newblock {Memory on the Racetrack}.
\newblock 10:195--198, 2015.

\bibitem{rthms}
I.~B. Peng~et al.
\newblock {RTHMS: A Tool for Data Placement on Hybrid Memory System}.
\newblock In {\em Proc. of the 2017 ACM SIGPLAN Int. Symposium on Memory
  Management}, ISMM 2017, pages 82--91, New York, NY, USA, 2017.

\bibitem{ArrayOrgDWM2016}
Z.~{Sun}, X.~{Bi}, W.~{Wu}, S.~{Yoo}, and H.~. {Li}.
\newblock {Array Organization and Data Management Exploration in Racetrack
  Memory}.
\newblock {\em IEEE Transactions on Computers}, 65(4):1041--1054, April 2016.

\bibitem{sun2013}
Z.~Sun, W.~Wu, and H.~Li.
\newblock {Cross-layer Racetrack Memory Design for Ultra High Density and Low
  Power Consumption}.
\newblock In {\em 50th ACM/IEEE Design Automation Conference (DAC)}, pages
  1--6, May 2013.

\bibitem{tapcache}
R.~Venkatesan~et al.
\newblock {TapeCache: A High Density, Energy Efficient Cache Based on Domain
  Wall Memory}.
\newblock In {\em Proceedings of the 2012 ACM/IEEE International Symposium on
  Low Power Electronics and Design}, ISLPED '12, pages 185--190, 2012.

\bibitem{xu2015}
H.~Xu, Y.~Li, R.~Melhem, and A.~K. Jones.
\newblock {Multilane Racetrack Caches: Improving Efficiency Through Compression
  and Independent Shifting}.
\newblock In {\em The 20th Asia and South Pacific Design Automation
  Conference}, pages 417--422, Jan 2015.

\end{thebibliography}

\end{document}